# Cash Transfers in the Perinatal Period and Child Welfare System Involvement Among Infants: Evidence from the Rx Kids Program in Flint, Michigan

November 24, 2025

Sumit Agarwal, MD, MPH, PhD[1]; H. Luke Shaefer, PhD[2]; Samiul Jubaed, MEd[2]; William Schneider, PhD[3]; Eric Finegood, PhD[4]; Mona Hanna, MD, MPH[4]

1. University of Michigan Medical School, School of Public Health, and Poverty Solutions, Ann Arbor, MI
2. Gerald R. Ford School of Public Policy and Poverty Solutions, University of Michigan, Ann Arbor, MI
3. University of Illinois at Urbana-Champaign School of Social Work, Urbana, IL
4. Michigan State University–Hurley Children's Hospital Pediatric Public Health Initiative, Charles Stewart Mott Department of Public Health, Michigan State University College of Human Medicine, Flint, MI

Corresponding Author: Sumit Agarwal, University of Michigan, 1500 E. Medical Center Drive, Ann Arbor, MI 48109. Email: sumitag@umich.edu.

Acknowledgements: The authors gratefully acknowledge Arianna Foster, Jenny LaChance, and Maya Wolock for their research assistance. The authors thank the Michigan Department of Health and Human Services and the Child and Adolescent Data Lab at the University of Michigan for assistance with data access. The authors also thank Alberto Abadie, John Ayanian, Nora Becker, Lawrence Berger, Lindsay Bullinger, Kyle Butts, Daniel Dench, Brian Jacob, Robert McClelland, Renuka Tipirneni, Sam Trejo, Atheendar Venkataramani, Jaume Vives-i-Bastida, and Jane Waldfogel for helpful comments. Finally, we are grateful to the City of Flint and the Rx Kids moms and their families. This research was supported by the Doris Duke Foundation and the Michigan Department of Health and Human Services. The contents are those of the authors and do not necessarily represent the official views of, nor an endorsement by, the Doris Duke Foundation and/or the Michigan Department of Health and Human Services.

# ABSTRACT

**Background:** Infants are most vulnerable to child maltreatment, which may be due in part to economic instability during the perinatal period. In 2024, Rx Kids was launched in Flint, Michigan, achieving near 100% aggregate take up and providing every expectant mother with unconditional cash transfers during pregnancy and infancy. This study examined the effect of Rx Kids implementation on child welfare system involvement.

**Methods:** This quasi-experimental study used administrative data from Children's Protective Services in Michigan, focusing on infants born in 2021 through 2024. The primary outcome was allegations of maltreatment within the first six months of life. Secondary outcomes included type of allegation (neglect versus non-neglect) and substantiated allegations. Synthetic difference-in-differences—a robust methodology for aggregate-level data and a single treated city—was used to compare changes in outcomes in Flint before and after implementation of Rx Kids relative to the corresponding change in control cities without the program.

**Results:** In the three years prior to the implementation of Rx Kids, the proportion of infants with a maltreatment allegation within the first six months of life was 21.7% in Flint and 19.5% among control cities. After implementation of Rx Kids in 2024, the maltreatment allegation rate dropped to 15.5% in Flint, falling below the maltreatment allegation rate of 20.6% among the control cities. Rx Kids was associated with a statistically significant 7.0 percentage-point decrease in the maltreatment allegation rate (p = 0.021), corresponding to a 32% decrease relative to the pre-intervention period. There was a decrease in the rate of neglect-related, non-neglect-related, and substantiated allegations; these were directionally consistent with the primary outcome but not statistically significant. Results were robust to alternative model specifications.

**Conclusions:** The Rx Kids prenatal and infant cash prescription program led to a significant reduction in allegations of maltreatment among infants. These findings provide important evidence about the role of economic stability in preventing child welfare system involvement.

**Introduction**

Child maltreatment, which includes child abuse and neglect, is a pernicious public health problem. Over one third of children in the U.S. experience an investigation for child maltreatment by their eighteenth birthday.[1] Children affected by maltreatment have worse mental and physical health later in life, are more likely to be involved in the criminal justice system, attain lower levels of education, and earn less into adulthood.[2–7] Each case of nonfatal child maltreatment is estimated to cost about $1.1 million in 2025 dollars, reaching a national economic burden of approximately $585 billion per year.[8] These costs extend beyond the immediate expenses, encompassing costs from health care, the child welfare and criminal justice systems, and other society-wide impacts.[7]

Poverty is a major risk factor for child maltreatment and involvement with the child welfare system.[9] Children living in low-income households experience maltreatment at 5 times the rate of other children and are 7 times more likely to experience neglect-specific maltreatment.[10] Living in communities of concentrated disadvantage also increases the risk for child maltreatment.[11,12] There are several possible interrelated mechanisms: Poverty increases the material hardships faced by families, including food and housing insecurity, and can limit parents' ability to purchase essential items for their children (e.g., diapers, car seats, adequate clothing, utilities). Poverty-related financial strain also contributes to family stress, which can undermine the capacity of parents to provide a safe, stable, and nurturing home environment, and increases parental risk of mental health challenges, substance use, partner conflict, and domestic violence (against children or an intimate partner).[13–16] Consequently, a growing body of research has found that income support programs in the U.S., such as refundable tax credits, can prevent child maltreatment.[17–34]

Infants (under 1 year of age) are the most vulnerable to maltreatment, with a substantiated allegation rate two to four times that of other pediatric age groups.[35] The heightened physical and mental demands of the perinatal period—coupled with the acute economic shock from decreasing household income and increasing expenses before and after birth—present unique vulnerabilities that can exacerbate the very risk factors associated with child maltreatment, including socioeconomic deprivation, parental substance use, stress and poor mental health, and housing instability.[36–39] Health-related factors in the perinatal period, such as inadequate prenatal care, low birthweight, and prematurity, also increase the risk for infant maltreatment.[40–42] Whether a time-bound cash transfer program, directed specifically at the perinatal period, can prevent allegations of maltreatment remains an open question and one of critical importance given the lifelong trajectories of early life adversity.

In January 2024, Rx Kids launched as a prescription for health, hope, and opportunity, in Flint, Michigan, a mid-sized city with one of the nation's highest child poverty rates. All expectant mothers residing in the city of Flint are eligible for a one-time transfer of $1,500 mid-pregnancy and $500 monthly after birth until age one. Rx Kids is the first cash transfer program in the U.S. that targets the perinatal period with universal eligibility (i.e., community-wide and no means testing based on income or asset thresholds). The program has achieved a high aggregate take-up of nearly 100%,[43,44] and prior work has demonstrated improved family financial security, prenatal care utilization, birth outcomes, and maternal mental health, leading to the hypothesis that the program could reduce child maltreatment and involvement with the child welfare system.[45–47] Our study tests this hypothesis by using a quasi-experimental study design to examine the effects of Rx Kids on the prevention of maltreatment allegations.

**Methods**

Data Sources

This study used administrative data from Children's Protective Services (CPS) in Michigan for all allegations of maltreatment among infants from January 2019 through June 2025. The data was provided to the University of Michigan's Child and Adolescent Data Lab under an agreement with the Michigan Department of Health and Human Services. The data included demographic information (e.g., date of birth, address), report date, details on the type of alleged maltreatment, and whether an allegation was substantiated. The data were geocoded based on address to the city level after which we used a deidentified version of the data for the analysis. In our primary analysis, we focused on allegations among infants born in 2021 through 2024 due to changes in reporting in 2019 and 2020 related to the Covid-19 pandemic.[48] Infants born in 2019 and 2020 were included in sensitivity analyses described below. The Michigan Department of Health and Human Services also provided the study team with statewide data on all births based on birth certificate records. We obtained additional data from the 2019-2023 U.S. Census Bureau's American Community Survey comprising the demographic and socioeconomic characteristics of the cities in which infants resided. This study was approved by University of Michigan's Institutional Review Board through the Child and Adolescent Data Lab.

Exposure

To address how poverty influences maternal and infant health, the Rx Kids cash prescription program was launched in Flint, Michigan, by Michigan State University-Hurley Children Hospital's Pediatric Public Health Initiative, in collaboration with Poverty Solutions at the University of Michigan. Rx Kids began enrolling participants in January 2024. Eligible participants enroll online and must provide documentation of 1) their residency within the city of Flint and 2) pregnancy or childbirth status as confirmed by medical documentation. After verification, the program provides participants with $1,500 during pregnancy (i.e., after 20 weeks gestation in 2024 and after 16 weeks gestation in 2025) and $500 per month for twelve months after birth, totaling $7,500 in unconditional cash and made available via direct deposit to a bank account or prepaid debit card. The program is administered by the nonprofit GiveDirectly and funded with support from the Temporary Assistance for Needy Families block grant, as well as other public and philanthropic grants. As a non-taxable gift to beneficiaries, the cash transfers are not subject to income taxes and do not affect eligibility for most other public benefits, including Medicaid and the Supplemental Nutrition Assistance Program.[49] No other major changes were identified in economic conditions or perinatal supports and services that would coincide with the timing of the implementation of Rx Kids in Flint. Given nearly 100% take up of Rx Kids as demonstrated in prior work,[43,44] the exposure was defined as Flint residence among infants born in 2024 after Rx Kids implementation.

Study Measures

The primary study outcome was maltreatment allegations within the first six months of life. Maltreatment allegations were aggregated to the annual level based on the infant's date of birth, and the maltreatment allegation rate was calculated using yearly denominators of city birth counts. Secondary outcomes included whether an allegation was substantiated after the investigation (substantiated allegations) and whether an allegation was classified as one related to neglect (neglect-related allegations) or non-neglect such as physical, emotional, or sexual abuse (non-neglect-related allegations).

Statistical Analysis

We used the quasi-experimental synthetic difference-in-differences method to compare changes in outcomes in Flint before and after implementation of Rx Kids relative to the corresponding change in cities without the program.[50,51] The method is the most appropriate and valid for situations such as ours with a single treated unit and group-level data among a panel of cities. The original synthetic control method was also considered for this analysis,[52,53] but as described further in the Appendix, this method is known to have problems when applied to short panels of data, which can lead to faulty statistical inference.[54,55] Synthetic difference-in-differences has been used in policy evaluations as a strategy for constructing a credible counterfactual; it does so by reweighting data from multiple untreated units and pre-intervention periods to approximate how the treated unit would have evolved absent the intervention.

Using synthetic difference-in-differences, the synthetic control for Flint was constructed from a donor pool of control cities without Rx Kids. Because the city of Flint has one of the highest child poverty rates (59% in 2023) and largest proportions of Black residents (56%) in the state—factors that are highly correlated with CPS reporting and involvement—we restricted the pool of potential control cities from elsewhere in Michigan based on pre-specified criteria used in prior Rx Kids studies.[46,47] The pre-specified criteria were mid-sized cities with a population of 5,000 to 125,000, a poverty rate of greater or equal to 15%, and a non-Hispanic Black population of at least 20%, which led to a donor pool of 21 control cities that resembled Flint most closely in terms of population size, poverty rate, and racial composition (see the Appendix for the full list of cities). Using data from the years prior to implementation of Rx Kids, the algorithm of synthetic difference-in-differences assigned weights to donor pool cities as well as time-specific weights to generate a synthetic control that follows the same trends in outcomes as Flint in the pre-intervention period and approximates how Flint would have evolved in the post-intervention period in the absence of Rx Kids. City characteristics, including poverty rate, median household income, racial and educational composition, additional household characteristics (proportion that were female-headed, renter-occupied, or housing-burdened), and total population were included as covariates. The coefficient of interest from synthetic difference-in-differences was estimated as the change in outcomes in Flint before and after the implementation of Rx Kids compared to the corresponding weighted change in control cities without the program.

Synthetic difference-in-differences combines attractive features of both the original synthetic control method and the difference-in-differences approach. Synthetic difference-in-differences is an extension of the synthetic control method that similarly reweights control cities,[56] but it differs from the synthetic control method in its use of both unit-specific and time-specific weights to achieve balance in pre-intervention outcomes, focusing on trends rather than absolute levels to establish parallel trends, a key assumption of the difference-in-differences framework for causal inference. Because it reweights in two dimensions, synthetic difference-in-differences possesses a form of double robustness, thus reducing the chance for bias from unobserved factors. In addition, its use of unit and time fixed effects absorbs additive differences across units and time, further strengthening its robustness relative to the original synthetic control method. Simulation results show that synthetic difference-in-differences dominates other estimators, including the synthetic control method.[50] These features of synthetic difference-in-differences improve statistical power, limit researcher degrees of freedom, and make it less sensitive to the composition of the donor pool and better suited to panels with only a few pre-intervention periods.

To examine whether the observed changes in Flint could have arisen due to chance, synthetic difference-in-differences use permutation-based placebo tests, a form of exact inference in which treatment is iteratively reassigned to control cities in the donor pool to generate a reference distribution against which a p-value can be calculated. All statistical tests were 2-tailed with a level of significance set at P<0.05. We conducted several sensitivity analyses to probe the robustness of our results to the exclusion of city characteristics as covariates, an alternative donor pool of control cities that was also available in the data consisting of 63 of the most populous cities in Michigan, and the length of the pre-intervention period. We also examined the results using the original synthetic control method, demonstrating how it leads to faulty inference in settings with a short panel. Finally, we assessed for changes in birth counts or any compositional change in who gives birth and conducted a sensitivity analysis focusing on the number of maltreatment allegations as an outcome. Additional details of our empirical strategy and sensitivity analyses are provided in the Appendix. Analyses were conducted using Stata version 19.5 (StataCorp).

**Results**

From January 2021 through June 2025, there were 40,332 allegations of child maltreatment within the first six months of life in Michigan among 404,292 infants born in 2021 through 2024 for a statewide maltreatment allegation rate of 10.0% over the study period. In the three years prior to the implementation of Rx Kids, the proportion of infants with a maltreatment allegation within the first six months of life was 21.7% (646/2971) in Flint and 19.5% (3921/20124) among control cities in the donor pool; the trend was relatively stable from 2021 to 2023 (Table 1). After implementation of Rx Kids in 2024, the maltreatment allegation rate dropped to 15.5% (165/1065) in Flint, falling below the maltreatment allegation rate of 20.6% (1303/6317) among the control cities (Figure 1). The maltreatment allegation rate in the control cities increased modestly over the same time period.

Using synthetic difference-in-differences to estimate the change in Flint before and after implementation of Rx Kids compared to the corresponding change in control cities without the program, Rx Kids was associated with a statistically significant 7.0 percentage-point decrease (95% CI, -12.9 to -1.0; p = 0.021) in the maltreatment allegation rate (Figure 2; Table 2), corresponding to a 32% decrease relative to the rate in Flint during the pre-intervention period. The weights used to generate the synthetic control are presented in Appendix Tables 1 and 2 and Appendix Figure 1. There was also a decrease in the rate of neglect-related, non-neglect-related, and substantiated allegations; these were directionally consistent with the primary outcome but not statistically significant.

In sensitivity analyses, the results were robust to excluding city characteristics as covariates, using an alternative donor pool of control cities consisting of 63 of the most populous cities in Michigan, and adding two additional years to the pre-intervention period (Table 3). Synthetic difference-in-differences was a more robust and appropriate method compared to the original synthetic control method (Appendix Table 3). As a final sensitivity check, we found a significant increase in the number of births in Flint but no significant compositional change (Appendix Table 4); furthermore, despite an increase in births in Flint, Rx Kids was associated with a significant reduction in the number of maltreatment allegations (-56.9 maltreatment allegations [95% CI, -76.4 to -37.3]).

**Discussion**

We found that the implementation of Rx Kids, the nation's first community-wide, unconditional prenatal and infant cash transfer program, was associated with a substantial and statistically significant population-level decline in the proportion of infants with a maltreatment allegation within the first six months of life. Prior to the implementation of Rx Kids, the maltreatment allegation rate in Flint was more than double the state rate (21.7% versus 10.0%). With the maltreatment allegation rate in Flint falling to 15.5% in 2024, this disparity between Flint and the state quickly narrowed by 50% within the first year of the program. Rx Kids prevented an estimated 57 infants from being involved in the child welfare system than would have been expected counterfactually without Rx Kids. These findings represent a consequential prevention of early life adversity with broad implications for children's health and wellbeing that could produce long-term savings and benefits for society.

These results are consistent with prior studies of income support programs showing that cash transfers can prevent child maltreatment.[17–34] Two particularly relevant studies examined such programs during the perinatal period. Rittenhouse (2023) used the discontinuity in tax benefits on January 1, around which infants born on either side of a calendar year receive very different refunds during infancy.[23] This study found that an additional $1,000 in tax benefits during infancy led to a 3% decline in the number of CPS referrals, investigations, and substantiated referrals by age three among low-income households. Bullinger et al. (2023) leverage year-to-year variation in cash transfers from the Alaska Permanent Fund Dividend.[24] The authors found that an additional $1,000 in dividend payments led to a 10% reduction in unsubstantiated allegations of child maltreatment and a 15% reduction in substantiated allegations by age 3. The magnitude of our estimates, scaled for the size of the Rx Kids cash transfer ($4,500 by age 6 months) and assuming the effects persist beyond six months of age, are closer to that of Bullinger et. al., possibly due to Rx Kids and the Alaska Permanent Fund Dividend both being universal in design and characterized by high take-up. Our findings alongside the prior literature support current policy efforts focused on providing concrete economic support to prevent child welfare system involvement, most notably embodied by the Family First Prevention Services Act of 2018.[57]

Rx Kids could have prevented involvement in the child welfare system through multiple, overlapping pathways. Poverty is a risk factor for, and increasingly recognized as causally linked to, child maltreatment, especially since neglect makes up 75% of all cases.[15] It may also be co-morbid with a range of factors that lead to neglect or abuse, including behavioral health conditions.[58] A prior study of Rx Kids has shown that the perinatal cash transfers were associated with reductions in food insecurity, housing instability, and other financial hardships.[45] In fact, Rx Kids led to a near-elimination of postpartum evictions among eligible mothers, consistent with evidence from eviction moratoria during the Covid-19 pandemic that reduced county-level rates of child maltreatment.[59] The Rx Kids program was also associated with improvements in maternal mental health and parenting stress.[45,60] Aligned with the family stress model, the reduction in financial strain and parental stress can in turn reduce the risk of child maltreatment.[61–63] Furthermore, premature and low birthweight infants are at the highest risk for maltreatment. Rx Kids reduced the rate of these adverse birth outcomes, suggesting yet another pathway that Rx Kids could have prevented involvement in the child welfare system by improving the perinatal environment.

This study has several limitations. First, because Rx Kids was not allocated randomly, there may be residual confounding unaccounted for by our quasi-experimental study design. The use of both unit and time weights establishes parallel trends to ensure the plausibility of the difference-in-differences strategy; it is analogous to techniques such as the adjustment of covariates to address the parallel trends assumption.[50] Second, our primary analysis focused on births between 2021 and 2024, a relatively short study duration with a single post-period year. The results, however, were robust to the inclusion of two additional years of pre-period data. Furthermore, trends were stable in the pre-intervention period, with each datapoint for Flint representing approximately 1000 births that limits the influence of random fluctuation. Third, we defined exposure as Flint residence among infants born in 2024 after Rx Kids implementation rather than actual enrollment in the program. Our analysis thus followed the intention-to-treat principle to capture the population-level effect of Rx Kids, a parameter that is of key interest to policymakers. In the setting of incomplete take up, the intention-to-treat estimates would normally be biased to the null compared to the effect of the program itself, but due to the high take-up rates of Rx Kids, our intention-to-treat estimates are likely to be very close to estimates for the effect of the program itself (i.e., treatment on the treated). Defining exposure based on residence also assumes that the city in which an infant resides at the time of an allegation six months into life is the same city in which the mother resided when the infant was born. This could produce some measurement error in our outcomes to the extent that there is migration in or out of communities. Fourth, there was evidence of a slight increase in the number of births without compositional change in Flint in 2024 compared to prior years, which could mechanically reduce the maltreatment allegation rate. Despite the increase in births, we found a substantial and significant decline in the absolute count of maltreatment allegations. Fourth, our analysis may not have been well-powered for the secondary outcomes in which we observed nonsignificant declines in the subcategories of maltreatment allegations. These were nonetheless directionally consistent with the primary outcome, and we follow the literature in focusing on maltreatment allegations as a principal outcome of interest.[64] Finally, the long-term implications of reducing allegations of child maltreatment during the first six months of life remain unknown, but preventing an initial allegation likely reduces the risk of future allegations,[65] consistent with the idea of improving lifelong trajectories starting from birth. Areas for future research include replication and persistence of findings after the first year of the program as well as in Rx Kids expansion communities for generalizability, examining long-term outcomes to age one and beyond, and further exploration of mechanisms of action.

This study found that Rx Kids, designed as a prenatal and infant cash prescription program to reduce perinatal poverty and improve health outcomes, yielded cross-sector benefits extending to the child welfare system. Infants are among the most vulnerable to child maltreatment, and Rx Kids was associated with substantial, population-level reductions in allegations of child maltreatment among infants in the earliest months of life. These findings underscore the importance of providing economic support during the perinatal period—a time of financial hardship and neurodevelopmental consequence—to improve the health and wellbeing of children.

**Figure 1: Maltreatment Allegations Within the First Six Months of Life for Infants in Flint, Control Cities, and State of Michigan**

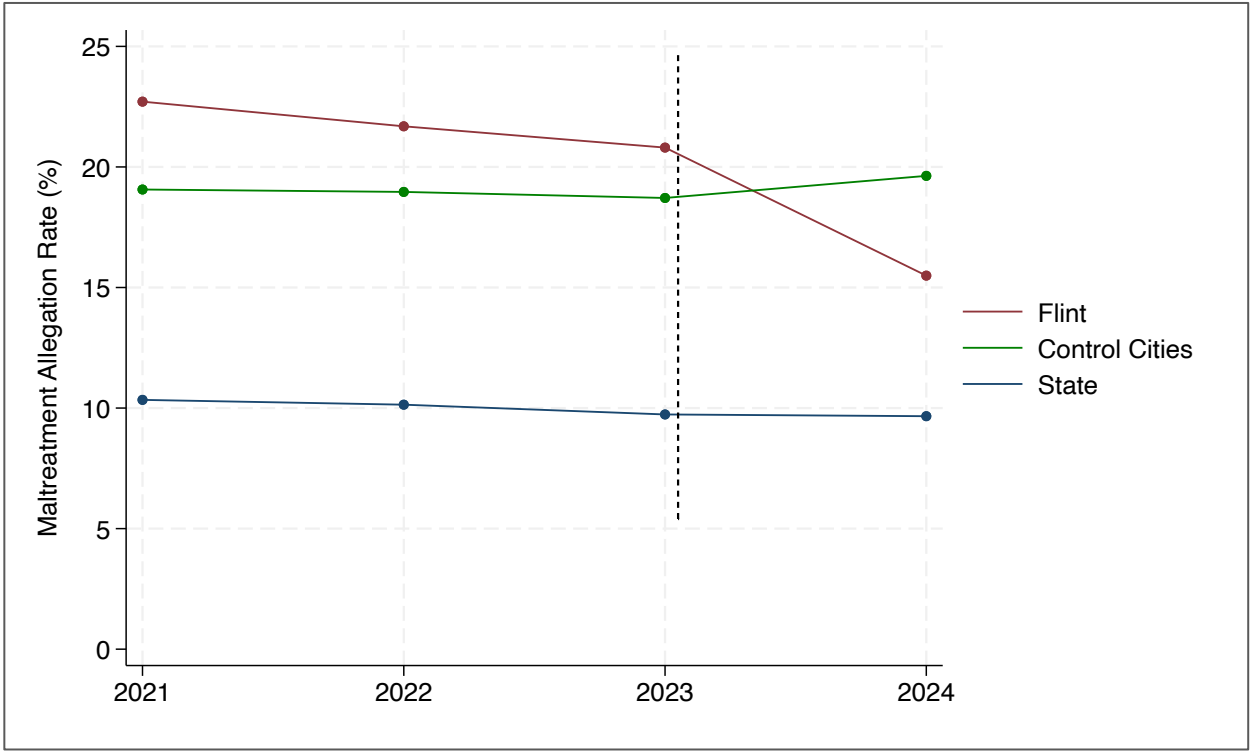

Note: The vertical dashed line differentiates the pre-intervention period (2021-2023) from the post-intervention period (2024).

**Figure 2: Synthetic Difference-in-Differences**

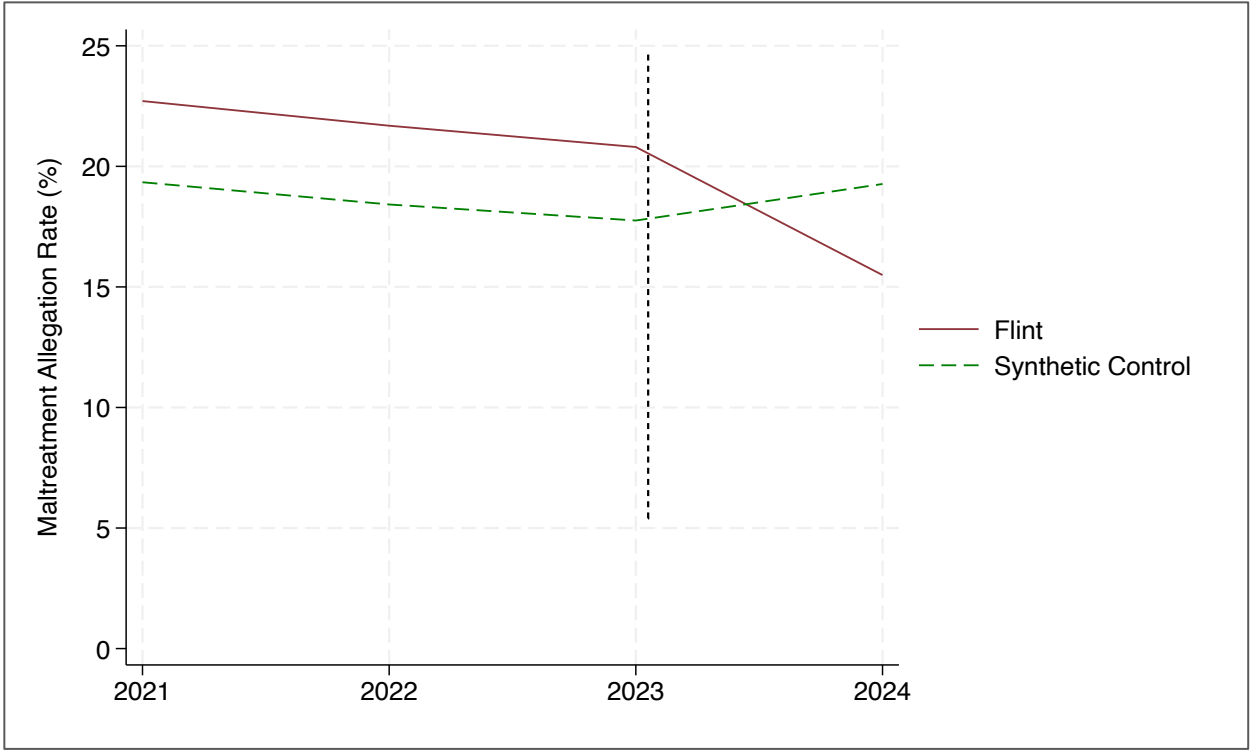

Note: The vertical dashed line differentiates the pre-intervention period (2021-2023) from the post-intervention period (2024).

**Table 1: Baseline Characteristics (Before Weighting)**

|  | Flint | Control Cities |
|---|---|---|
| Maltreatment allegations, % | 21.7 | 18.9 |
|     Maltreatment allegations (2021), % | 22.7 | 19.1 |
|     Maltreatment allegations (2022), % | 21.7 | 19.0 |
|     Maltreatment allegations (2023), % | 20.8 | 18.7 |
| Substantiated allegations, % | 3.0 | 4.5 |
| Poverty rate, % | 34.4 | 26.7 |
| Deep poverty rate, % | 18.1 | 12.4 |
| Median household income, $ | 36,194 | 44,552 |
| Racial composition, % | | |
|   Hispanic | 4.5 | 9.0 |
|   Non-Hispanic Black | 56 | 45.4 |
|   Non-Hispanic White | 32.6 | 38.7 |
|   American Indian and Alaska Native | 0.3 | 0.3 |
|   Asian | 0.6 | 1.1 |
|   Pacific Islander | 0.0 | 0.0 |
|   Other | 0.5 | 0.4 |
|   Multiple Races | 5.5 | 5.1 |
| Educational composition, % | | |
|     Less than high school diploma, % | 16.7 | 14.1 |
|     At least high school diploma, % | 83.3 | 85.9 |
| Households with children and female head of household, % | 59.3 | 53.4 |
| Households that are renter-occupied, % | 45 | 43.8 |
| Households that are housing-burdened, % | 54.4 | 50.5 |
| Gini index | 0.48 | 0.46 |
| Public health insurance, % | 70.5 | 56.1 |
| Total population | 80,835 | 25,933 |

**Table 2: Estimates from Synthetic Difference-in-Differences for the Primary and Secondary Outcomes**

| Outcome | Mean Rate in Flint Prior to 2024 (%) | Synthetic Difference-in-Differences | |
|---|---|---|---|
| | | Estimate (95% CI) | $P$ value |
| Maltreatment allegations | 21.7 | -7.0 pp (-12.9 to -1.0) | 0.021 |
|     Neglect-related | 10.0 | -3.6 pp (-8.2 to 1.0) | 0.13 |
|     Non-neglect-related | 11.7 | -3.4 pp (-9.0 to 2.3) | 0.24 |
| Substantiated allegations | 4.5 | -1.3 pp (-4.2 to 1.6) | 0.36 |

Note: All estimates are interpreted as a percentage-point (pp) change.

**Table 3: Sensitivity Analyses**

|  | Maltreatment Allegations | |
| --- | --- | --- |
| **Specifications** | **Estimate (95% CI)** | ***P* value** |
| *Primary study period (2021-2024)* | | |
|    Pre-specified donor pool | -7.0 pp<br>(-12.9 to -1.0) | 0.021 |
|    Alternative donor pool | -5.8 pp<br>(-10.0 to -1.5) | 0.008 |
| *Adding two additional years to the pre-intervention period (2019-2024)* | | |
|    Pre-specified donor pool | -6.3 pp<br>(-11.7 to -0.9) | 0.022 |
|    Alternative donor pool | -7.9 pp<br>(-12.5 to -3.4) | 0.001 |

Note: All estimates are interpreted as a percentage-point (pp) change. The pre-specified donor pool is comprised of 21 control cities that resembled Flint most closely in terms of population size, poverty rate, and racial composition. The alternative donor pool is comprised of 63 of the most populous cities in Michigan.

**Appendix**

<u>Table of Contents</u>

A. Supplementary Methods

B. Supplementary Tables and Figures

C. Supplementary References

**A. Supplementary Methods**

The criteria for inclusion in the donor pool of control cities from elsewhere in Michigan included cities with a population of 5,000 to 125,000, a poverty rate of greater or equal to 15%, and a non-Hispanic Black population of at least 20%, as determined using the U.S. Census Bureau's 2019-2023 American Community Survey. After excluding areas adjacent to the city of Flint such as Beecher and Flint Township as well as Kalamazoo which had its own perinatal programming during our study period, the twenty-one cities that met the criteria included Albion, Benton Harbor, Benton Township, Bridgeport Township, Buena Vista Township, Eastpointe, Ecorse, Harper Woods, Highland Park, Inkster, Jackson, Lansing, Muskegon, Muskegon Heights, Pontiac, River Rouge, Saginaw, St. Louis (in Michigan), Wayne, Ypsilanti, and Ypsilanti Township.

A potential concern with this donor pool, despite being pre-specified, is that it includes a relatively small number of cities as well as a few relatively small-sized cities. We conducted a sensitivity analysis using a distinct donor pool of control cities that was also available in the data, specifically 63 of the most populous cities in Michigan excluding townships and similarly excluding Kalamazoo as above. These 63 cities included: Adrian, Allen Park, Ann Arbor, Auburn Hills, Battle Creek, Bay City, Birmingham, Burton, Canton, Clinton, Dearborn, Dearborn Heights, Detroit, East Lansing, Eastpointe, Farmington Hills, Ferndale, Flint, Forest Hills, Garden City, Grand Rapids, Hamtramck, Holland, Holt, Inkster, Jackson, Kentwood, Lansing, Lincoln Park, Livonia, Madison Heights, Marquette, Midland, Monroe, Mount Pleasant, Muskegon, Norton Shores, Novi, Oak Park, Okemos, Pontiac, Port Huron, Portage, Redford, Rochester Hills, Romulus, Roseville, Royal Oak, Saginaw, Saint Clair Shores, Shelby, Southfield, Southgate, Sterling Heights, Taylor, Troy, Walker, Warren, Waterford, Waverly,

Westland, Wyandotte, Wyoming, and Ypsilanti. This was not our preferred donor pool specification because the criterion of the largest cities diverged from the pre-specified criteria used in prior Rx Kids studies, and it leads to a donor pool substantially different than Flint on key dimensions that are related to child maltreatment. For example, the median poverty rate of these 63 cities was 11.6% (which is different than Flint by almost a factor of three), and the median proportion of non-Hispanic Black residents was 8.2% (which is different than Flint by nearly a factor of seven). This is particularly important when using the synthetic control method, as discussed in further detail in Appendix Table 3. The features of synthetic difference-in-differences (see main text) make it less sensitive to the composition of the donor pool.

**Appendix Table 1: Unit Weights for the Primary Outcome (Maltreatment Allegations) in the Synthetic Difference-in-Differences Analysis**

| Donor City | Weights |
|---|---|
| Albion | 0.04651482 |
| Benton Harbor | 0.07667844 |
| Benton Township | 0.03837185 |
| Bridgeport Township | 0.05365189 |
| Buena Vista Township | 0.04815389 |
| Eastpointe | 0.05666218 |
| Ecorse | 0.0606078 |
| Harper Woods | 0.04317521 |
| Highland Park | 0.02471664 |
| Inkster | 0.0159125 |
| Jackson | 0.04504911 |
| Lansing | 0.04650577 |
| Muskegon | 0.03825902 |
| Muskegon Heights | 0.03533298 |
| Pontiac | 0.03159653 |
| River Rouge | 0.07526972 |
| Saginaw | 0.05019247 |
| St. Louis | 0.06235585 |
| Wayne | 0.04630271 |
| Ypsilanti | 0.04773139 |
| Ypsilanti Township | 0.05695925 |

**Appendix Table 2: Time Weights for the Primary Outcome (Maltreatment Allegations) in the Synthetic Difference-in-Differences Analysis**

| Year | Time Weights |
| --- | --- |
| 2021 | 0.43497877 |
| 2022 | 0.14212496 |
| 2023 | 0.42289627 |

**Appendix Table 3: Results from the Original Synthetic Control Method with a Short Panel**

| Specification | Maltreatment Allegations | | | |
| --- | --- | --- | --- | --- |
| | Estimate | RMSPE Ratio for Flint | Range of RMSPE Ratios from Placebo Tests | P value |
| *Primary study period (2021-2024)* | | | | |
| Pre-specified donor pool (21 cities) | -6.9 pp | 800 billion | 0.29 to 5.35 | 0 |
| Alternative donor pool (63 cities) | -7.1 pp | 38 billion | 0.27 to 4.48 trillion | 0.38 |
| *Adding two additional years to the pre-intervention period (2019-2024)* | | | | |
| Pre-specified donor pool (21 cities) | -7.4 pp | 7.78 | 0.17 to 3.27 | 0 |
| Alternative donor pool (63 cities) | -7.7 pp | 6.59 | 0.31 to 519 billion | 0.46 |

Notes: All estimates are interpreted as a percentage-point (pp) change. The results using the original synthetic control method show effect sizes that are similar across all specifications. Flint had the largest or second largest reduction in the maltreatment allegation rate compared to the placebo tests for all other units in the pre-specified and alternative donor pools. Statistical inference under the synthetic control method, however, is highly variable depending on the number of years in the pre-intervention period and the number of cities in the donor pool. In the synthetic control method, p-values are determined by comparing the root mean square prediction error (RMSPE) ratio for Flint against a reference distribution of RMSPE ratios from placebo tests (Abadie, Diamond, and Hainmueller 2010, 2015; Abadie 2021). Such high RMSPE ratios into the billions are clear evidence of overfitting to idiosyncratic noise, the risk of which increases with short panels of data and large donor pools (Abadie and Vives-i-Bastida 2022; Hollingsworth and Wing 2020). The synthetic control method is not appropriate or valid in such a setting. Extending the pre-intervention period, when possible, and trimming the donor pool to units similar to the treated unit in the pre-intervention period can correct this issue of overfitting, as suggested by the 3rd specification listed in the table that used 5 years in the pre-intervention period and a smaller group of donor units more similar to Flint (i.e., the 21 control cities in the prespecified donor pool that resembled Flint most closely in terms of population size, poverty rate, and racial composition). Synthetic difference-in-differences has several features that improve upon known limitations of the synthetic control method (see main text) and was thus our primary method for studying the effect of Rx Kids on child welfare system involvement.

**Appendix Table 4: Results from Synthetic Difference-in-Differences for Composition of Births, Number of Births, and Number of Allegations**

| Outcome | Synthetic Difference-in-Differences | |
|---|---|---|
| | Estimate (95% CI) | *P* value |
| Maternal age: 16-26 | -1.6 pp (-12.6 pp to 9.3) | 0.77 |
| Non-Hispanic Black race | 3.3 pp (-9.9 to 16.5) | 0.62 |
| Maternal education: less than high school | 1.0 pp (-9.6 to 11.7) | 0.85 |
| Not married | 2.4 pp (-5.6 to 10.4) | 0.56 |
| Medicaid as payor | 3.5 pp (-15.9 to 22.9) | 0.72 |
| First birth (i.e., parity of 0) | 1.9 pp (-8.2 to 12.0) | 0.71 |
| Number of births | 96.7 (35.0 to 158.4) | 0.002 |
| Number of allegations | -56.9 (-76.4 to -37.3) | < 0.001 |

Notes: There was no statistically significant change in the composition of who gives birth using synthetic difference-in-differences. Furthermore, the point estimates for several of these would, if anything, suggest higher risk of child maltreatment allegations in the post-period from any potential compositional change. As shown in prior work (reference 47), there was similarly no statistically significant change in the composition of who gives birth using individual-level data from birth certificates and a difference-in-differences approach. There was a statistically significant increase in the number of births in Flint; despite an increase in births in Flint, Rx Kids was associated with a significant reduction in the number of maltreatment allegations.

**Appendix Figure 1: Pre-Intervention Balance Across Control Units from Synthetic Difference-in-Differences for the Primary Outcome (Maltreatment Allegations)**

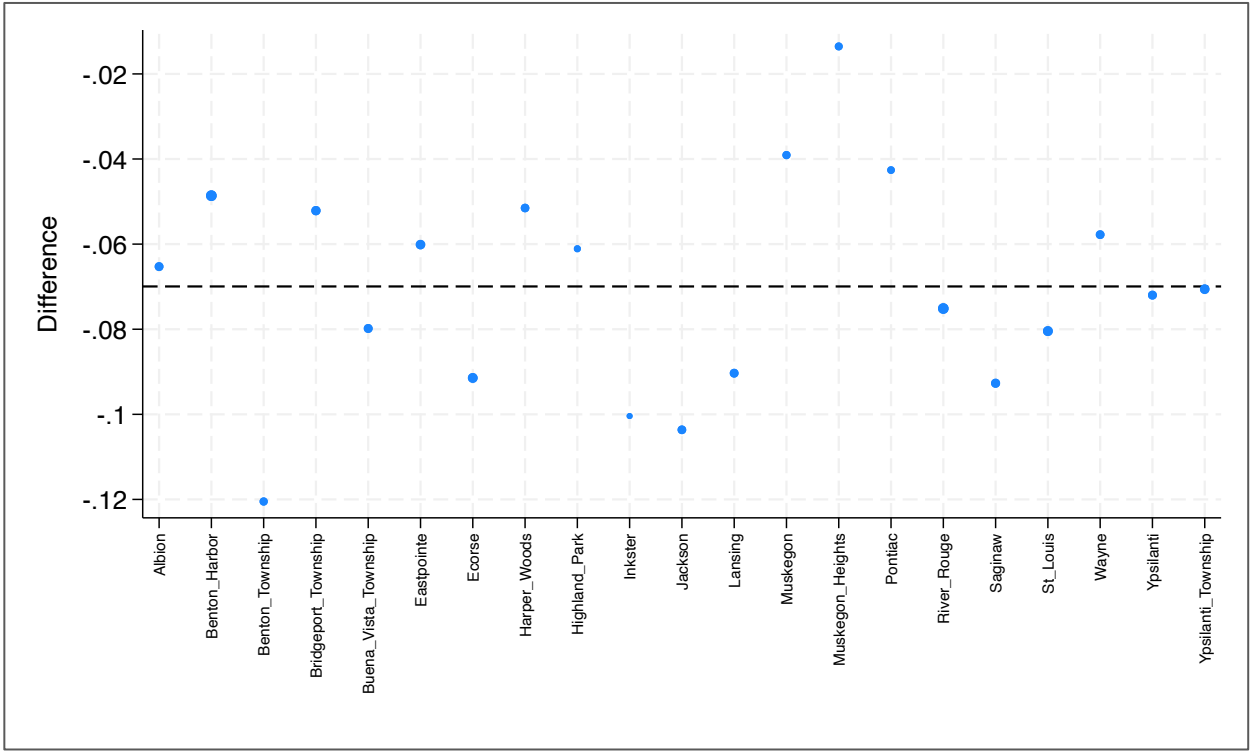

Note: The raw unit and time weights from synthetic difference-in-differences are shown in Appendix Table 1 and 2. This plot shows how well each control unit's pre-intervention outcomes align with Flint after optimizing the time weights.

**Supplementary References**